# Voltage Controlled Spin-Orbit Torque Switching in W/CoFeB/MgO

Jinsong Xu[*] and C.L. Chien[#]

Department of Physics and Astronomy, Johns Hopkins University, Baltimore, Maryland 21218, USA

**Abstract**

Voltage control of magnetism and spintronics have been highly desirable, but rarely realized. In this work, we show voltage-controlled spin-orbit torque (SOT) switching in W/CoFeB/MgO films with perpendicular magnetic anisotropy (PMA) with voltage administered through $SrTiO_3$ with a high dielectric constant. We show that a DC voltage can significantly lower PMA by 45%, reduce switching current by 23%, and increase the damping-like torque as revealed by the first and second-harmonic measurements. These are characteristics that are prerequisites for voltage-controlled and voltage-select SOT switching spintronic devices.

Authors to whom correspondence should be addressed:

[*] jxu94@jhu.edu and [#] clchien@jhu.edu

Electrical control of magnetism and spin phenomena without using a magnetic field has been a long-standing goal in spintronics. The discovery of spin-transfer-torque (STT) earlier accommodates electrical switching of ferromagnetic (FM) entities with spin-polarized currents [1,2]. More recently, the advent of spin-orbit torque (SOT) via Rashba effect and spin Hall effect (SHE) using pure spin current has been demonstrated for switching magnetization in ferromagnet(FM)/heavy metal(HM) bilayers with geometries that are not possible in conventional STT devices [3-8]. A pure spin current has the distinct attributes of delivering spin angular momentum with a minimal number of charge carriers in metals and no charge carriers in insulator. With different heavy metals (HMs), SOTs of various magnitudes and sign have achieved in switching FM entities with in-plane anisotropy, and also perpendicular magnetic anisotropy (PMA), if additional measure is taken to break the up/down symmetry.

It is of great interest to explore voltage reduction of SOT switching current and especially to achieve voltage controlled switching so that only the specific device among many can be selected by voltage to execute switching on demand [9-11]. It has been shown that voltage can modify the magnetic anisotropy of some FM layers via inducing surface charges [12-14] or modifying the orbit occupancy at the oxide/metal interface [15-19]. As a result of the modification of interface anisotropy or exchange bias by electric field, electric-field-assisted magnetization switching has been reported in magnetic tunnel junctions [20-25], as well as modulation of spin Hall nano-oscillator [26-28]. There have been some attempts in voltage-control SOT switching [29-33], but with very small effects and obscure underlying physics.

In this work, we report efficient voltage control of SOT switching in the well-known W/CoFeB/MgO thin films with PMA, where the gate voltage is applied via $SrTiO_3$ with a high dielectric constant. The coercivity field ($H_c$) and critical current for switching ($I_c$) are reduced by 45% and 23%, respectively, by a gate voltage of a few volts. The gate voltage not only affects anisotropy as alluded above, but also the efficiency of SOT switching via the pure spin current as revealed by the first and second harmonic measurements. Only a gate voltage of a suitable sign results in reduced magnetic anisotropy field ($H_k$) but increased damping-like SOT effective field ($H_{DL}$), leading to different reductions in $H_c$ and $I_c$.

We pattern Hall bar devices for this study as schematically shown in Fig. 1(a), where the W (2)/CoFeB(1)/MgO(1.5)/TaO$_x$(1) multilayered films (where the numbers in parentheses are thicknesses in nm) have been made by sputtering at room temperature on Si wafer with 300 nm SiO$_2$ layer. To achieve PMA, the as-grown films were annealed in an out-of-plane field of 1T at 280 °C in vacuum for 30 min. We used photolithography and ion milling to fabricate the Hall bar structure with 8-μm wide channels. A SrTiO$_3$ layer 200 nm in thickness was sputtered onto the device as the dielectric layer at 77 K to preserve PMA, before the final Au top layer. More information about the device fabrication can be found in the Supplementary Material Note 1. The gate voltage is applied between the top Au electrode and the CoFeB layer to generate an electric field. The high dielectric constant of SrTiO$_3$ enables large electric field without leakage current [10,34]. The suitable range of applied gate voltage varies from -2 V to 20 V, beyond this range either the devices break down or the PMA compromised.

A single CoFeB layer has in-plane anisotropy (IPA) because of the large demagnetization factor $4\pi M_s$, where $M_s$ is the magnetization. However, when sandwiched between a HM and a suitable oxide, W/CoFeB/MgO acquires surface anisotropy of $-4K_s/M_s t$, where $K_s$ is the surface anisotropy constant. With a positive $K_s$, at a sufficiently small thickness $t$, the surface anisotropy alters the total anisotropy from IPA to PMA as in our sample of W (2)/CoFeB(1)/MgO(1.5) [35]. One may exploit the delicate PMA with a gate voltage and achieve voltage-control switching.

We use anomalous Hall effect (AHE) to demonstrate PMA in the W/CoFeB/MgO multilayers. As shown in Fig. 1 (b), there is a sharp square AHE hysteresis loop with an out-of-plane magnetic field $\mu_0 H_z$, confirming PMA in the W/CoFeB/MgO heterostructures. Under a gate voltage, the width of the AHE hysteresis loop varies as shown by the red (at -2 V) and blue (at 20 V) curves in Fig. 1 (b), revealing the systematic modification of the coercivity field by the gate voltage as shown in Fig. 1 (c). The coercivity $\mu_0 H_c$ monotonically decreases from 5.8 mT to 3.2 mT when varying $V_{gate}$ from -2 V to 20 V, aided by the large dielectric constant of $SrTiO_3$. In general, the coercivity field depends on both the magnetic anisotropy and microstructure of the materials [36,37]. Here the low gate voltage changes mainly the magnetic anisotropy. Previous studies have shown that electric field can modify the orbital hybridization between the Fe *3d* orbitals and the O *2p* orbitals to alter magnetic surface anisotropy [15-19]. The unidirectional change of $H_c$ with both negative and positive gate voltages is consistent with this scheme. As we will discuss later, the gate voltage affects $H_k$ only by about 10% but resulting in a much larger change of coercivity $H_c$ by 45%.

This modification of the coercivity by a gate voltage enables voltage controlled magnetic field switching. For example, when a 4.2 mT $\mu_0 H_{pulse}$ field pulse is applied opposite to the magnetization, the magnetization of CoFeB is readily switchable at $V_{gate} = 20$ V but remains unchanged at -2 V (see Supplementary Material Note 2) because the 4.2 mT $\mu_0 H_{pulse}$ is larger than $\mu_0 H_c$ at 20 V but smaller than $\mu_0 H_c$ at -2 V.

In addition to voltage controlled magnetic field switching, we explore the prospect of gate voltage on SOT switching. As it is well known that because the spin index of the pure spin current is parallel to the interface, the SOT is incapable of switching a PMA layer without some means to break the up/down symmetry, such as a magnetic field is also applied along the current direction. The SOT switching behavior of W/CoFeB/MgO at $V_{gate} = -2$ V and 20 V with an external field of 20 mT along the current direction is shown in Fig, 2a. We note there are asymmetry and intermediate states of the SOT switching hysteresis loops for one current sweeping direction. These are likely due to the different processes of domain wall depinning and expansion in the Hall cross between the two current sweeping directions [38]. The width of the SOT switching loop at $V_{gate} = 20$ V is narrower than that at $V_{gate} = -2$ V, indicating the gate tunability of the critical current $I_c$ required to switch magnetization ($I_c$ is defined as the current required to switch $R_{xy}$ to 0). The gate voltage dependence of $I_c$ is shown in Fig. 2(b), where $I_c$ decreases from 0.73 mA ($3.0 \times 10^{10}$ A m$^{-2}$) to 0.56 mA ($2.3 \times 10^{10}$ A m$^{-2}$) by ~23% when varying $V_{gate}$ from -2 V to 20 V. The monotonic decrease of $I_c$, even when $V_{gate}$ changes sign, illustrates that it is an electric field effect influencing the charge transfer at the CoFeB/MgO interface. The modification of switching current by a gate voltage clearly demonstrates voltage-controlled SOT switching. For example, when a current pulse $I_{pulse}$ of magnitude 0.75 mA ($3.1 \times 10^{10}$ A m$^{-2}$) of the appropriate polarity is applied to W/CoFeB/MgO, the magnetization of CoFeB is readily switchable at $V_{gate} = 20$ V but not at -2 V (see Supplementary Material Note 3).

There are two terms of SOT in a HM/FM/oxide heterostructure, $a\mathbf{M} \times \boldsymbol{\sigma} + b\mathbf{M} \times (\boldsymbol{\sigma} \times \mathbf{M})$, known as the field-like torque and the damping-like torque, respectively. The effect of the two terms can be viewed as the field-like effective field $\mathbf{H_{FL}} \sim \boldsymbol{\sigma}$ and the damping-like effective field $\mathbf{H_{DL}} \sim \boldsymbol{\sigma} \times \mathbf{M}$. In general, the SOT switching efficiency depends on the intrinsic magnetic properties of the FM layer and $I_c \propto M_s H_k / H_{DL}$ [3,39]. To determine the gate voltage dependence of the two SOT terms, we perform first and second harmonic measurements. Referring to Fig. 1a, an AC excitation current $I_{ac}$ with a magnitude of 0.3 mA ($1.25 \times 10^{10}$ A m$^{-2}$) is applied along the *x* direction, the transverse in-phase first harmonic ($V_{1\omega}$) and out-of-phase second harmonic ($V_{2\omega}$) voltages along the *y* direction are simultaneously measured by two lock-in amplifiers [40]. Fig. 3(a) shows the manner $V_{1\omega}$ evolves as an external magnetic field $H_x$ is applied along $x$ direction (actual magnetic field is applied with a small out-of-plane tilting angle to ensure coherent domain wall rotation). The value of $V_{1\omega}$ decreases faster as a function of $H_x$ at $V_{gate}$ = 20 V (blue circles) than that at $V_{gate}$ = -2 V (red circles), indicating a weaker magnetic anisotropy at a larger gate voltage. To quantify $H_k$, we follow the Stoner-Wohlfarth model,

$$V_{1\omega} = I_{ac}R_{xy} \propto R_{AHE}\frac{M_z}{M_s} = R_{AHE}\frac{H_k+H_z}{\sqrt{(H_k+H_z)^2+H_x^2}} \qquad (1)$$

where $R_{AHE}$ is the anomalous Hall coefficient, with the extracted values of $H_k$ and $R_{AHE}$ shown in Fig. 3(b) and Fig. 3(d). Across the gate voltage range of -2V to 20 V, $H_k$ changes by about 10%, which contributes to the reduction of $I_c$ because of the aforementioned linear dependence. This modification of $H_k$ by gate voltage is usually attributed to the change of $d$ states electron density of CoFeB [16-19,41], which appears to be responsible for the small change of $R_{AHE}$ as a result of the Fermi level change. We have also observed similar behavior when the magnetic field is applied along $y$ direction (shown in Supplementary Material Note 4). Based on the gate dependent $R_{AHE}$ vs. $H_x$ curves, the voltage control magnetic anisotropy coefficient is estimated to be about 30 fJ V$^{-1}$ m$^{-1}$ (see Supplementary Material Note 5), which is similar to previous reports [23,42].

The SOT effective field $H_{DL}$ and $H_{FL}$ can be obtained by [3,40,43]

$$H_{DL(FL)} = -2\frac{B_{x(y)} \pm 2\xi B_{y(x)}}{1-4\xi^2} \qquad (2)$$

where the $\pm$ sign corresponds to the $\pm M_z$ state, $B_{x(y)} \equiv \frac{\partial V_{2\omega}}{\partial H_{x(y)}}/\frac{\partial^2 V_{1\omega}}{\partial H_{x(y)}^2}$, $\xi = \frac{R_{PHE}}{R_{AHE}}$ is the ratio of planar Hall effect (PHE) coefficient $R_{PHE}$ to the anomalous Hall coefficient $R_{AHE}$. To extract $R_{PHE}$, we apply a constant in-plane magnetic field $\mu_0 H_{ext}$ = 600 mT, which is larger than the perpendicular anisotropy field $\mu_0 H_k$ ~200 mT, and sweep the angle $\varphi$ between current direction and magnetic field direction [shown in Fig. 3(c)]. In this configuration, as the magnetization is nearly in-plane, $R_{xy}$ mainly comes from PHE, which is proportional to $sin2\varphi$. There is also a small contribution from AHE caused by the unintentional small out-of-plane titling angle $\theta_0$, which is proportional to $sin\varphi$. Therefore, we fit the data using $R_{\text{PHE}}sin2\varphi + R_{\text{AHE}}sin\theta_0 sin\varphi$ (solid lines in Fig. 3(c)). We find $R_{\text{PHE}}$ is about 1 Ω, which is much smaller than the previously measured $R_{\text{AHE}}$ (~10 Ω), therefore $\xi = \frac{R_{\text{PHE}}}{R_{\text{AHE}}}$ is about 0.1 in the applied gate voltage range.

The second-harmonic measurement results are shown in Fig. 4 (a). When the magnetic field applied along the current direction (*x* direction), the slopes of $V_{2\omega}$ as a function of field are the same for both magnetization states $\pm M_z$ [Fig. 4(a) upper panel], while they have opposite sign when the field is applied perpendicular to the current direction (*y* direction) [Fig. 4(a) lower panel]. The solid curves are the results of linear fit to the data. At a larger gate voltage (blue curves), $V_{2\omega}$ has larger slope with respect to the magnetic field, which provides a hint that there might be a larger SOT. By combining the results of first and second harmonic measurements and planar Hall measurements, we can calculate the SOT quantitatively, and the results are shown in Fig. 4(b). Because of the abovementioned inverse proportionality between $H_{\text{DL}}$ and $I_c$, we focus on the damping-like SOT effective field $H_{\text{DL}}$ (there is some variation of field-like torque at different gate voltages, but no clear trend is observed, see Supplementary Material Note 6). As seen in Fig. 4(b), $H_{\text{DL}}$ increases with $V_{\text{gate}}$ by about 10% in the applied gate voltage range, that causes the reduction of $I_c$. This modification of $H_{\text{DL}}$ by gate voltage may result from the alteration of interfacial Rashba effect [30], spin mixing conductance and/or the change of effective spin Hall angle, which depends on the interfacial electronic structure and scattering rates influenced by the electric field [29,32,44,45]. Together with the 10% decrease of the magnetic anisotropy $H_k$, they can account for the ~23% reduction of $I_c$. We have observed similar behaviors in other devices (see

Supplementary Material Note 7). Therefore, a gate voltage can tune both the magnetic anisotropy and SOT, resulting in the modification of the critical current for switching.

In conclusion, we demonstrate voltage control of field switching and SOT switching in perpendicularly magnetized W/CoFeB/MgO multilayers, facilitated by the usage of SrTiO$_3$ with a large dielectric constant. The coercivity and critical current are reduced by 45% and 23% respectively, with a gate voltage under 20 V. The harmonics measurements show that the reduction of critical current originates from both the alteration of magnetic anisotropy and damping-like SOT effective field. The tunability of coercivity and critical current by a gate voltage demonstrate the prospects of voltage-controlled switching for future energy efficient memory devices.

See Supplementary Material for detailed description of device fabrication, voltage controlled magnetic field switching, voltage controlled spin-orbit torque switching, first harmonic measurement with $H_y$ field, estimation of voltage control magnetic anisotropy coefficient, gate dependence of field-like torque and additional data from a second device.

Authors to whom correspondence should be addressed:

*E-mail: jxu94@jhu.edu and [#]E-mail: clchien@jhu.edu

**Acknowledgement**

We acknowledge the technical assistance from Yufan Li. Work at JHU was supported by the U.S. Department of Energy, Basic Energy Science Award No. DE-SC0009390.

**Data Availability**

The data that support the findings of this study are available from the corresponding authors upon reasonable request.

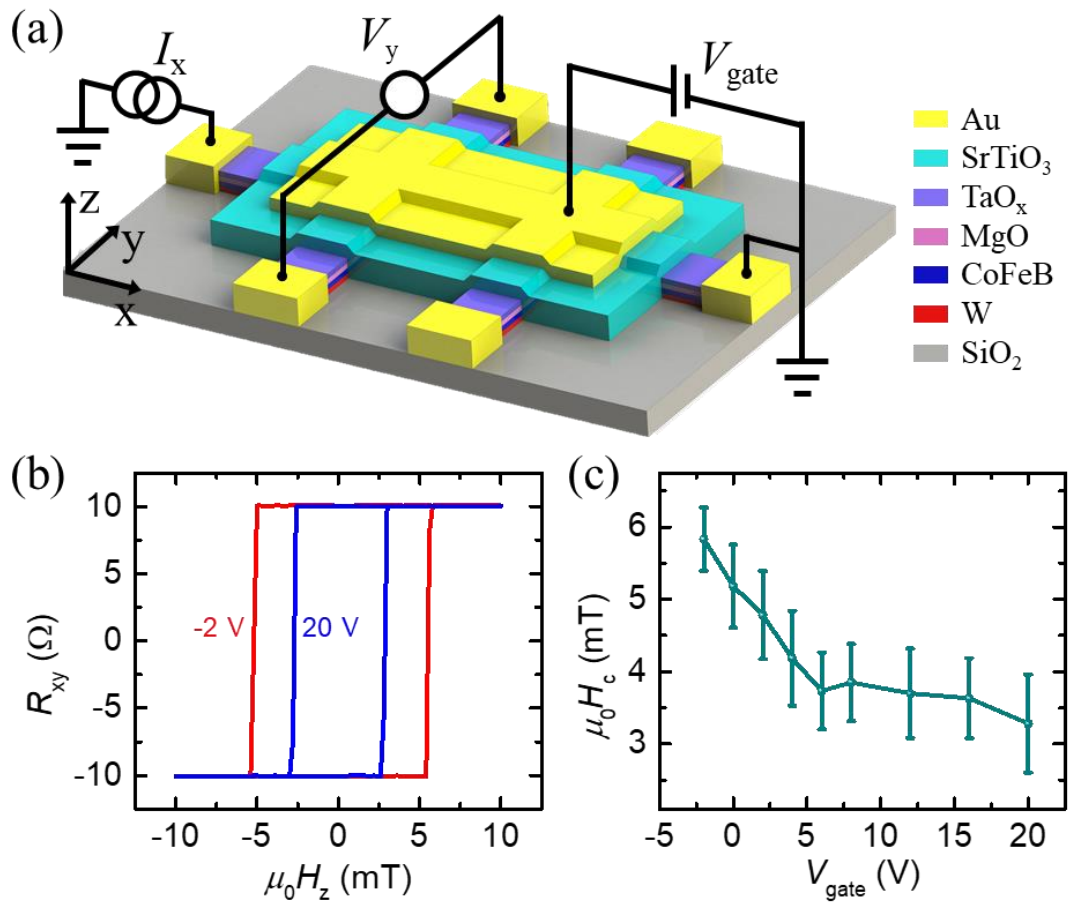

FIG. 1. (a) Schematic of the Hall bar device with layer order shown on the right. (b) Anomalous Hall effect measured at $V_{gate}$ = -2 V (red) and 20 V (blue) with an out-of-plane magnetic field. (c) Gate voltage dependence of coercivity field.

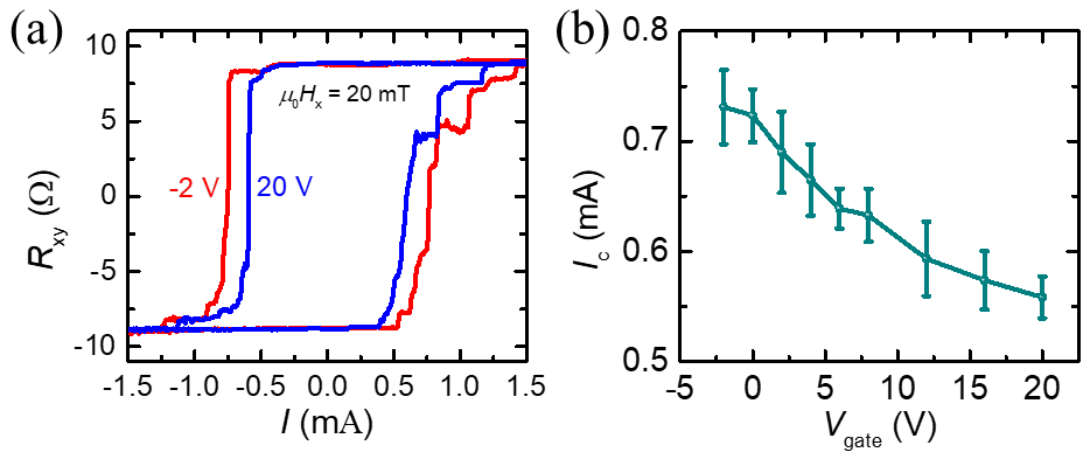

FIG. 2. (a) SOT induced magnetization switching at $V_{gate}$ = -2 V (red) and 20 V (blue) with an assisted field of $\mu_0 H_x$ = 20 mT. (b) Gate voltage dependence of critical current $I_c$ required to switch magnetization.

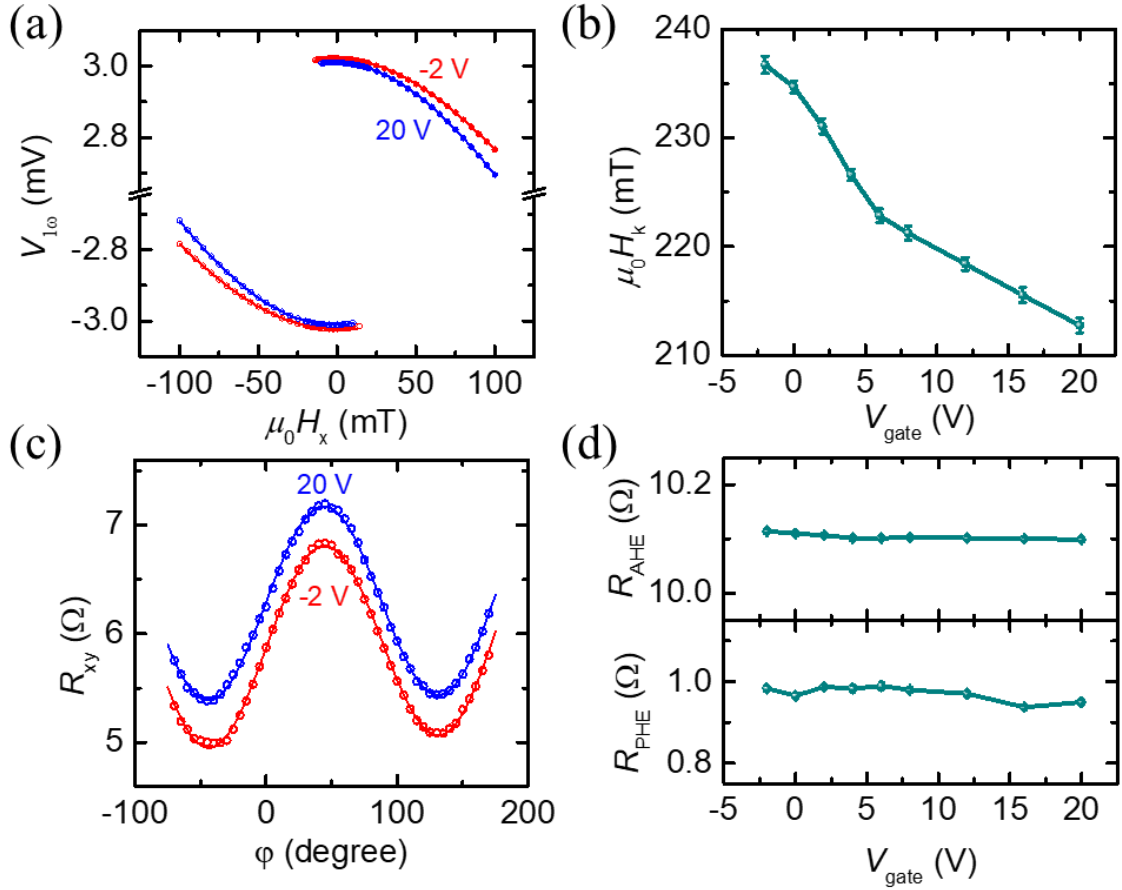

FIG. 3. (a) First harmonic measurements at $V_{gate}$ = -2 V (red) and 20 V (blue). Solid (open) circles represent magnetization along + (-) $z$ direction. The solid curves are best-fit to the data using Eq. (1) with an AC excitation current of 0.3 mA. (b) Gate voltage dependence of magnetic anisotropy $\mu_0 H_k$, extracted from first harmonic data using Eq. (1). (c) $R_{xy}$ at $V_{gate}$ = -2 V (red) and 20 V (blue) as a function of in-plane angle φ at constant in-plane magnetic field $\mu_0 H_{ext}$ = 600 mT with the solid curves as the best-fits. (d) Gate voltage dependence of anomalous Hall coefficient $R_{AHE}$ and planar Hall coefficient $R_{PHE}$, extracted from (a) and (c), respectively.

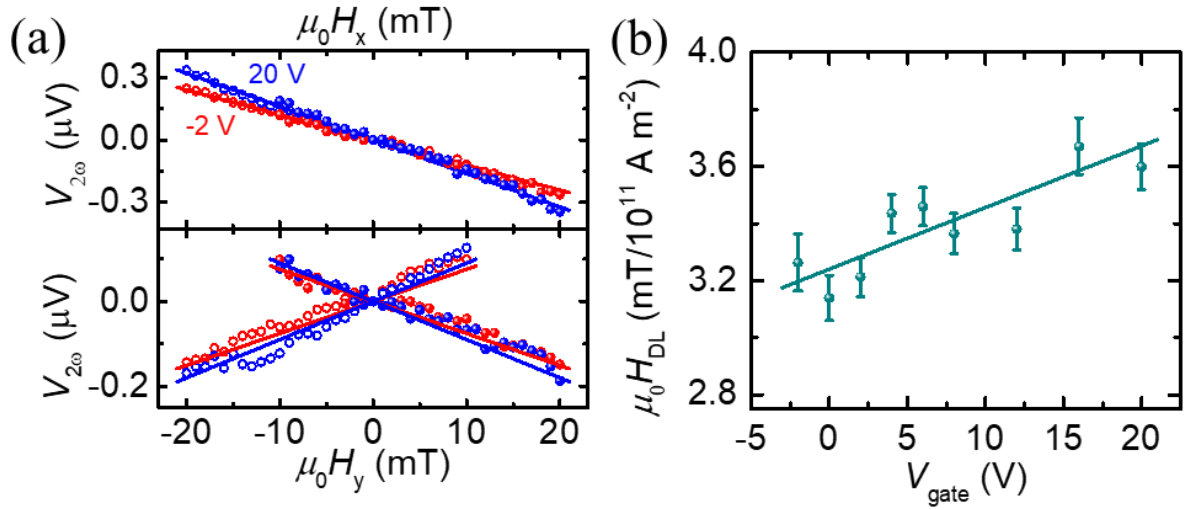

FIG. 4. (a) Second harmonic measurement with magnetic field applied along *x* (upper panel) and *y* direction (lower panel) at $V_{gate}$ = -2 V (red) and 20 V (blue). Solid (open) circles represent magnetization pointing along + (-) *z* direction. The solid lines are the linear fits to the data. (b) Gate voltage dependence of damping-like SOT effective field $\mu_0 H_{DL}$, extracted from first and second harmonic measurements with the AC excitation current at 0.3 mA ($1.25 \times 10^{10}$ A m$^{-2}$). The solid line is the linear fit to the data.